\documentclass[12pt]{article}
\usepackage{amsmath,amssymb,graphicx}
\usepackage{enumerate}
\usepackage{cite}
\def\be{\begin{equation}}
\def\ee{\end{equation}}
\def\ai{\'{\i}}
\def\lp{\left(}
\def\rp{\right)}

\begin{document}
\title{\vspace{-1.5cm} \bf Tides and energy conditions in Einstein--Gauss--Bonnet thin-shell wormholes}

\author{Ernesto F. Eiroa$^{1,}$\thanks{e-mail: eiroa@iafe.uba.ar}, Emilio Rub\'{\i}n de Celis $^{2,3,}$\thanks{e-mail: erdec@df.uba.ar},  
Claudio Simeone$^{2,}$\thanks{e-mail: csimeone@df.uba.ar}\\
{\small $^1$ Instituto de Astronom\'{\i}a y F\'{\i}sica del Espacio (IAFE, CONICET-UBA),} \\ 
{\small  Ciudad Universitaria, 1428, Buenos Aires, Argentina}\\
{\small $^2$ Instituto de F\'{\i}sica de Buenos Aires (IFIBA, CONICET-UBA),}\\
{\small Ciudad Universitaria Pabell\'on I, 1428, Buenos Aires, Argentina}\\ 
{\small $^3$ Departamento de F\'{\i}sica, Facultad de Ciencias Exactas y Naturales,} \\ 
{\small Universidad de Buenos Aires, Ciudad Universitaria Pabell\'on I, 1428, Buenos Aires, Argentina}}

\date{\small \today}

\maketitle
\vspace{0.6cm} 
\begin{abstract}

In this article we study spherical thin-shell wormholes in five-dimensional Einstein--Gauss--Bonnet gravity. We show that configurations supported by non-exotic matter, that is matter satisfying the weak energy condition, are possible at the same time that traversability problems associated with strong radial tides at the throat can be avoided when suitable values of the parameters are adopted. Our construction is performed in such a way that it also allows for the admissible behaviour of the geometry in the whole spacetime.

\end{abstract}

\section{Introduction}

Wormholes are spacetime geometries in which two regions (of the same universe or of two universes) are connected by a throat, that is a surface where geodesics open up \cite{motho,book}. This means a non trivial topology, which is associated to notable features as, for example, the possibility of time travel --with all the resulting questions implied (see \cite{mty88} and \cite{frno90}). Within the framework of general relativity all compact wormhole configurations require to be threaded by exotic matter, that is matter which does not satisfy the energy conditions \cite{hovis1}. In thin-shell wormholes \cite{vis89} such matter is localized in a layer placed at the wormhole throat. This has the advantage of keeping within spatial bounds the matter with not very desirable properties, and it also allows for a straightforward analysis of the mechanical stability of the matter distributions under perturbations preserving the symmetry \cite{poivis95,ernro04,ern08,mglv12}. Other aspect for which these wormhole models are well suited is the study of thermodynamic stability \cite{fmh19,ern24}. However, the thin-shell approach  introduces a possible additional difficulty: a singular matter distribution generates a discontinuity in the derivatives of the spacetime  metric, and this means that large tides could appear jeopardizing the actual --in a practical sense-- traversability across the wormhole. We have already addressed this problem as well as other closely related ones in our papers \cite{nos21,nos22,nos23,nos25}.

A possible way to avoid the problems with matter and tides can be to work in the wider framework of gravity theories beyond relativity, in which the relation between matter and geometry differs from that in Einstein equations. For example, it was shown that in pure Gauss--Bonnet gravity \cite{gra-wi} wormholes could exist even with no matter (see also \cite{7,maeda,gggw08}). Moreover, in our  paper \cite{risi08} we found 5-dimensional Einstein--Maxwell--Gauss--Bonnet spherical thin-shell wormholes supported by ordinary matter, which were later proven to be stable under mechanical perturbations preserving the symmetry \cite{maz}. In a subsequent work we showed that, starting from a different branch of Wiltshire spherically symmetric solution \cite{wilt}, wormhole geometries could exist without exotic matter even for small positive values of the Gauss--Bonnet coupling constant (that is, little departures from pure relativity) and with both the electric charge and the cosmological constant equal to zero\cite{yo11} (which substantially differs from the case studied here; see below). More recently, in our article \cite{egc19} in the same framework it was shown that charged but massless and asymptotically flat wormholes were possible.

Here we address the question if, enabled by the different relation between matter and geometry given by the Gauss--Bonnet contribution to the equations of gravitation, the interesting situation is possible in which a thin-shell wormhole is supported by non exotic matter,  and tides across its throat are acceptable for an extended body; this is not possible within pure Einstein gravity. This implies an analysis of the Riemann tensor --in terms of which tides are expressed-- associated to singular matter distributions, as well as to examine the junction conditions --relating the surface energy-momentum on the shells with the curvature jumps-- extended to Einstein--Gauss--Bonnet theory \cite{davis4}. We will mainly restrict the study to positive values of the Gauss--Bonnet parameter, and this will compel us to work with the so called non-GR branch (or ``Gauss--Bonnet branch'') of Wiltshire solution (see below). We will be able to show the existence of configurations with the desired properties; besides, possible problems with the behaviour of the metric associated to the required values of the cosmological constant will be avoided by suitable conditions imposed on the mathematical cut and paste construction of the complete geometry.

\section{The geometry}

 The Einstein--Gauss--Bonnet equations of gravitation in five spacetime dimensions have the form
\begin{equation}
R_{\mu\nu}-\frac{1}{2}g_{\mu\nu}R+ \Lambda g_{\mu\nu}+2\alpha H_{\mu\nu}=8\pi G_5T_{\mu\nu}\label{field}
\end{equation}
where $\Lambda$ is the cosmological constant, $G_5$ is the Newton constant in five dimensions (we set $G_5=1$ in the units used here) and $\alpha$ is the constant of dimension $(\rm{length})^2$ multiplying the Gauss--Bonnet terms 
\begin{equation} 
H_{\mu\nu}=RR_{\mu\nu}-2R_{\mu\alpha}R^{\alpha}_{\nu}-2R^{\alpha\beta}R_{\mu \alpha\nu\beta}+R_{\mu}^{\alpha\beta\gamma}R_{\nu\alpha\beta\gamma}-\frac{1}{4}g_{\mu\nu}(R^{2}-4R^{\alpha\beta}R_{\alpha\beta}+R^{\alpha\beta\gamma\delta}R_{\alpha\beta\gamma\delta})
\end{equation}
where $R_{\mu\nu}$ and $R_{\mu \alpha\nu\beta}$ are, respectively, the Ricci and Riemann tensors. The equations of Einstein--Gauss--Bonnet gravity coupled to Maxwell electromagnetism admit spherically symmetric solutions of the form  \cite{wilt}
\begin{equation}
{ds}^{2}=-f(r)\,dt^{2}+f^{-1}(r)\,dr^{2}+r^{2}d\Omega^{2}_{3}\label{metric0}
\end{equation}
where $d\Omega^{2}_{3}$ is the line element of a unitary 3-sphere and
\begin{equation}
f(r)=1+\frac{r^{2}}{4\alpha}\left[1\mp\sqrt{1+\frac{16M\alpha}{\pi r^{4}}-\frac{8Q^{2}\alpha}{3 r^{6}}+\frac{4 \Lambda \alpha}{3}}\right]\label{metric}
\end{equation}
with $M$ the ADM mass and  $Q$ the electric charge. The $(-)$ sign corresponds to the so-called GR branch of the solution, 
because, for $\Lambda =0$, it reproduces  for large $r$ the five dimensional Reissner--Nordstr\"{o}m solution (or the Schwarzschild one if the charge is zero); in the same limit the $(+)$ solution, called non-GR branch\footnote{The $(+)$ solution is also called the ``exotic'' branch, while the $(-)$ solution is also known as the ``normal'' one. Because here we will use ``normal'' and ``exotic'' for matter satisfying or not the energy conditions, we will avoid such
 denominations.}, acquires a sort of effective cosmological constant $\Lambda_\mathrm{eff}\propto -\alpha^{-1}$ and its mass term has the ``wrong" sign,
 which implies a repulsive force on rest particles (recall the signs in the solution within $5D$ Einstein gravity; see, for instance, \cite{ec5d}).  A central feature of the non-GR solution is the absence of horizons and then the existence of a naked singularity at the origin. However, in a wormhole construction in which a minimum radius exists (the throat radius $r=a>0$) this is admissible. Besides, an advantage regarding the matter content of the shell placed at the throat will be apparent when we write down the components of the surface energy-momentum tensor (see \cite{yo11,egc19}). 

A static wormhole geometry can be defined by removing the regions $r<a$ in two copies of the geometry given by (\ref{metric0}) and (\ref{metric}) and matching the two submanifolds with $r\geq a$  at a static spherical surface $\Sigma$ of  radius $a$. The symmetry across $\Sigma$ ensures the continuity of the line element. The flare-out condition to be dealing with a properly defined wormhole, that is that geodesics open up at the surface $\Sigma$, is automatically verified because of the simple $r^2$ dependence of the angular components of the metric. The matching conditions relating the geometries at the two sides  with the character of matter on this surface (the energy-momentum tensor of the shell placed at $\Sigma$) depend on the way in which the Gauss--Bonnet terms are regarded: if these terms are understood as a sort of additional contribution to the energy-momentum tensor, they are the usual ones known as    Darmois--Israel conditions \cite{sen,lanc,darm,isr}; the matter content and stability of wormholes constructed within that framework was studied in \cite{grg06}. If, instead, the Gauss--Bonnet terms are regarded as an essentially geometric object, as we do throughout the present work, the junction conditions are rather different, as found in  \cite {davis4} starting from the equations above; they read
\begin{equation}
\langle K_{ij}-K h_{ij}\rangle + 2\alpha \langle 3J_{ij}-Jh_{ij}+2P_{iklj}K^{kl}\rangle =-8\pi S_{ij},\label{field-shell}
\end{equation}
where  $\langle Y \rangle$ stands for  the jump $Y^+-Y^-$ of a given quantity across the  surface $\Sigma$, $h_{ij}$ is the surface induced metric, $S_{ij}$ is the surface energy-momentum tensor, and $K_{ij}$ are the components of the extrinsic curvature which read
 \be
K_{ij}^\pm = - n_\gamma^\pm \left. \left( \frac{\partial^2
  X^\gamma}{\partial \xi^i \partial \xi^j} + \Gamma_{\alpha \beta}^\gamma
  \frac{\partial X^\alpha}{\partial \xi^i} \frac{\partial
  X^\beta}{\partial \xi^j} \right) \right|_{r=a},
\ee
where $n_\gamma^\pm$ are the unit normals to the surface $\Sigma$ at each side of it, $\xi_i$ are the
coordinates on this surface, and $X_\mu$ are the coordinates of the $(4+1)$-dimensional manifold; the divergence-free part of the Riemann tensor $P_{ijkl}$ and the tensor $J_{ij}$ are defined as follows:
\begin{equation}
P_{ijkl}=R_{ijkl}+(R_{jk}h_{li}- R_{jl}h_{ki})-(R_{ik}h_{lj}- R_{il}h_{kj})+\frac{1}{2}R(h_{ik}h_{lj}-h_{il}h_{kj}),
\end{equation}
\begin{equation}
J_{ij}=\frac{1}{3}\left[2KK_{ik}K^{k}_{j}+K_{kl}K^{kl}K_{ij}-2K_{ik}K^{kl}K_{lj}-K^{2}K_{ij}\right]
\end{equation}
(see \cite{davis4} and also \cite{mae}). Note that Eq. (\ref{field-shell}) includes three extra terms with respect to general relativity; when $\alpha =0$ this equation reduces to the corresponding one in the Darmois--Israel formalism\footnote{On the other hand, in the particular case of having no matter at the joining surface (i.e. $S_{ij}=0$), which is typical in the study of stars, it was recently shown \cite{brassel} that the proper matching in EGB gravity reduces to only require the continuity of both the surface induced metric and the extrinsic curvature, as it also results in the same scenario in general relativity.}. For a static configuration, the explicit components of the extrinsic curvature for a metric of the form (\ref{metric0}) are
\be
{K_{\theta_k }^{\theta_k }}^\pm =\frac{1}{a}\sqrt{f_\pm(a)},
\ee
\be
{K_{t}^{t}}^\pm =  \frac{f'_\pm(a)}{2\sqrt{f_\pm(a)}}\ ,\label{e9} 
\ee
where $\theta_k $ are the angular coordinates and a prime stands for a derivative with respect to $r$. The resulting expression for the surface energy density on the shell is then
\begin{equation}
\sigma(a)=-S_t^t=-\frac{\sqrt{f(a)}}{8\pi a}\left[ 6-\frac{4\alpha}{a^2}\left(2f(a)-6\right)\right]\label{sigma}
\end{equation}
and the isotropic pressure has the form
\begin{equation}
p(a)=S_{\theta_k}^{\theta_k}=\frac{\sqrt{f(a)}}{8\pi a}\left[ 4+\frac{af^{'}(a)}{f(a)}+4\alpha  \frac{ f^{'}(a)}{af(a)}\left(1-f(a)\right)\right].\label{pe}
\end{equation}
In the general relativity limit $\alpha=0$ the energy density is negative, which is consistent with the well known requirement of exotic matter supporting compact wormholes in the framework of Einstein's gravity \cite{hovis1,book} (non compact configurations, as cylindrically symmetric wormholes, admit an alternative definition of the flare-out condition, which in turn is compatible with a positive energy density; see \cite{brle09}).

\section{Tides}

Tides are described in terms of  the covariant relative acceleration $(\Delta {\cal A})^\mu$, which is proportional to the Riemann tensor ${R^\mu}_{\alpha\nu\beta}$  (see  \cite{grav,book}). If we consider an oriented small separation $(\Delta x)^\nu$ between two points of a body moving with four-velocity $V^\alpha$, we have
\be\label{A0}
(\Delta {\cal A})^\mu =
- {R}^\mu_{\alpha\nu\beta} V^\alpha (\Delta x)^\nu V^\beta.
\ee   
The problem of tides in geometries associated to singular distributions of matter  is essentially the issue with jumps of the curvature in the joining surfaces where the matter is located; no aspects of particular interest arise associated to the bulk contributions in the nearby smooth regions \cite{martin,genc}. The treatment for such configurations starts from carefully expressing all tensors taking into account the jumps involved\footnote{See for instance Chapter 14 of \cite{book}.}; the detailed calculation can be found in \cite{nos21}, and also in \cite{nos22,nos23}. We begin by recalling the result for the radial tide
on a radially moving object: the covariant relative acceleration between two  points of an object traversing a spherical wormhole throat is 
\be\label{A}
\Delta {\cal A}_r=-
{\kappa^{t}}_{t}+\frac{\Delta \tilde\eta}{2} \left[  {{R^{rt}}_{rt}}^{-} +{{R^{rt}}_{rt}}^{+}\right]_{a} 
+ \mathcal{O}(\Delta \tilde\eta^2) 
,
\ee
where ${\kappa^{t}}_t$ is the jump of the component ${K^{t}}_t$ of the extrinsic curvature across the infinitely thin shell, and $\Delta\tilde\eta$ is the proper radial separation between the points considered; this reduces to $\Delta\eta$ for a  body at rest. The nearby smooth regions contribute with a term given by the components of the Riemann tensor at each side of the shell; this is proportional to $\Delta\tilde\eta$, and describes a tension which vanishes in the limit $\Delta\tilde\eta\to 0$. The jump in the extrinsic curvature, instead, adds a finite non zero contribution which we note $
{\cal T}_r$, and is fixed for a given geometry: it does not vanish for infinitely close points at different sides of the throat (besides, this implies not a tension but a compression). This particular nature of tides reflects the discontinuous character of the gravitational field (i. e. of the first derivatives of the metric) at {$r=a$}; it implies that  wormhole geometries supported by infinitely thin matter layers, though traversable in principle, present the practical problem of possible great tides acting on a body extended across the throat. Because for a static wormhole symmetric across the throat we have
\begin{equation}
{\kappa^{t}}_t=\frac{f^{'}(a)}{\sqrt{f(a)}},\label{T}
\end{equation}
the possibility of ${\cal T}_r=0$, that is to avoid traversability issues for a radially extended object traversing the wormhole throat in a radial trajectory \cite{nos21,nos22,nos23}, implies $f'(a)=0$.

In an analogous way, following \cite{nos21} we decompose the angular tide on a radially moving object as a finite part plus a divergent one:
 \be \label{D_a_perp_gen}
\Delta {\cal A}_\perp
= \Delta {\cal A}_\perp^{\mathrm{finite}} + \Delta {\cal A}_\perp^{\mathrm{div}}.
\ee
Given the spherical symmetry, the result must be the same for any angular coordinate; then to simplify the notation we choose the angle $\varphi$, and we can write
\begin{eqnarray}
\Delta {\cal A}_\perp^{\mathrm{finite}} & = &
\frac{\Delta x_{\perp}}{2} \,
\left[
{{R^{\varphi t}}_{\varphi t}}^-
+\gamma^2 \beta^2 \,
 \left(
 {{R^{\varphi t }}_{\varphi t}}^- -  \, {{R^{\varphi r}}_{ \varphi r}}^-
\right)
\right]_{a}\nonumber\\
& & +\,\frac{\Delta x_{\perp}}{2}\left[{{R^{\varphi t}}_{\varphi t}}^+ +\gamma^2 \beta^2 \,
 \left(
 {{R^{\varphi t }}_{\varphi t}}^+ -  \, {{R^{\varphi r}}_{ \varphi r}}^+
\right)
\right]_{a}.
\end{eqnarray}
On the other hand, introducing the infinitely-short travelling proper time $\delta\tau$ of the object across the shell we have 
\be
\Delta {\cal A}_\perp^{\mathrm{div}}  
=
 \frac{1}{\delta\tau}\Delta x_\perp\,\gamma \beta \,
{\kappa^{\varphi}}_{\varphi}\,.\label{aperp}
\ee
Here, as usual, we  define the parameters $\gamma = 1/\sqrt{1-\beta^2}$ and the positive radial speed $\beta$ of the object as measured in an orthonormal frame. The divergent character of the result is apparent because of the proper time $\delta\tau\to 0$ for an infinitely thin shell. Differing from the case of radial tides, for tides in transverse directions to the radial one both contributions are proportional to the transverse extension $\Delta x_{\perp}$ of the object. For an angular tide, the finite term implies a compression produced by the curvature of the smooth parts of the geometry, while the divergent term corresponds to a stretching effect resulting from the jump in the extrinsic curvature associated to the flare-out at the shell. The possibility of straightforwardly cancelling the divergence by a certain choice of the parameters is ruled out by the existence of a throat, that is a surface where a flare-out condition is satisfied --or in other words, where geodesics open up: this excludes ${\kappa^{\varphi}}_{\varphi}=0$. Transverse tides would be strictly zero only for a  body at rest (see (\ref{aperp})), while they could be reduced by traversing the throat with a low speed, this in the understanding that the crossing time is not zero for a realistic shell with little but not vanishing thickness $\epsilon$. In such case we can write
\be
\Delta {\cal A}_\perp^{\mathrm{(div)}}  
=
\frac{1}{\epsilon}\Delta x_\perp\, \gamma^2 \beta^2 \,
{\kappa^{\varphi}}_{\varphi}\,,\label{aperp2}
\ee
where we have writen (div) to note the component that would be divergent in the zero thickness limit. Because for the class of geometries studied here the angular components of the extrinsic curvature jump are given as
\be
{\kappa^{\varphi}}_{\varphi}=2\frac{\sqrt{f(a)}}{a},
\ee
the traversing speed could be tuned according to the shell thickness, as well as all other parameter relations satisfying  energy and other traversability conditions (see below), to achieve admissible values of the transverse tide on an extended body. The situation with the radial tide is essentially different: though not formally divergent, it cannot be controlled by such kind of adjustments even if a non zero thickness shell is admitted. The radial tide for two points at different sides of the shell does not depend of the separation between these points, nor on the traversing speed of the body, but is it essentially given by twice the radial acceleration towards the center just at each side, and is therefore fixed by the jump of the extrinsic curvature (see the discussion in Sec. 2 of \cite{nos21}). This motivates the following analysis.

\section{Energy conditions and traversability}

We want to examine the possibility of achieving the interesting situation in which a thin-shell wormhole would be supported by normal matter and would be acceptable in what regards to radial tides on an extended object traversing the throat. Ordinary matter satisfies the so called weak energy condition\footnote{We are assuming an isotropic configuration and matter as a type-I non composite fluid, so the form of the energy condition is the usual one; see \cite{maemar2020} and the related discussion in  \cite{bras2}.} $\sigma\geq 0$, $\sigma +p \geq 0$.
 Expressions (\ref{sigma}) and (\ref{pe}) lead to 
\begin{equation}
\sigma (a)+p(a)=-a\sigma'(a)/3\label{s+p}.
\end{equation}
Then to fulfil the condition $\sigma(a)+p(a)\geq 0$ we could ask for $\sigma '(a)\leq 0$. However, if our aim is to achieve a null contribution to radial crossing tides coming from the --ideally-- singular nature of the matter distribution, we can start working from the beginning under the restriction ${\cal T}_r=0$ which forces  $f'(a)=0$. This simplifies the expression for the pressure which, besides, is then forced to be positive and independent of the Gauss--Bonnet parameter:
\be 
p(a)|_{{\cal T}_r=0}=\frac{\sqrt{f(a)}}{2\pi a}>0.
\ee
As an important consequence, $\sigma(a)\geq 0$ implies $\sigma(a) +p(a)>0$ and  ordinary matter on the shell at the wormhole throat is ensured just by a non negative energy density. 

\subsection{Conditions at the wormhole throat}

The non trivial form of the metric function $f$ and the number of parameters involved allow to succeed in our aim; nevertheless, other problems appear as a result of the peculiar behaviour of $f_+(r)$, even if no horizons must be dealt with. If we introduce the definitions
\be\label{defs}
\frac{16M\alpha}{\pi a^{4}}\equiv \psi,\ \ \  \ \ \frac{4 \Lambda \alpha}{3}\equiv \chi, \ \ \ \  \ -\frac{8Q^{2}\alpha}{3 a^{6}}\equiv \zeta
\ee
then for the GR branch of the solution given by (\ref{metric}) the condition $\sigma\geq 0$ implies
\begin{equation}
-8\alpha-2a^2-a^2\sqrt{1+\psi+\chi+\zeta}\geq 0
\end{equation}
which imposes $\alpha<0$ (see \cite{risi08} and also \cite{gggw08}); in this case a non vanishing charge would help to reduce the required absolute value of the Gauss--Bonnet parameter. Instead, for the non-GR branch the energy condition $\sigma\geq 0$ yields
\begin{equation}
-8\alpha-2a^2+a^2\sqrt{1+\psi+\chi+\zeta}\geq 0.\label{s>0}
\end{equation}
This is possible for the more desirable theoretical framework in which the Gauss--Bonnet parameter is positive: in the derivation of the theory as a low energy limit of heterotic string theory (see \cite{bode85,zwi85,grwit86,grsl87}), $\alpha$ is proportional to the inverse string tension, and should then be positive\footnote{If such alternative derivation is not adopted, there are still other reasons to assume $\alpha>0$, as shown for example in \cite{yco22} in relation with the consistency of holographic descriptions.}. This implies that we must have  
\be
\sqrt{1+\psi+\chi+\zeta}>2. \label{cond}
\ee
Note that in this case the sign of the electric charge contribution is such that a non vanishing $Q$ makes more difficult to fulfil the positive energy condition. From now on we  restrict our analysis to positive values of the Gauss--Bonnet parameter and we will deal, unless explicitly noted otherwise, with the non-GR branch of Wiltshire solution given by $f_+$. 

In what regards in itself the condition ${\cal T}_r=0$ of no radial tides associated to the singular nature of the matter distribution, ${\cal T}_r\propto f_+'(a)=0$ determines that
\be
1+\sqrt{1+\psi+\chi+\zeta}=\frac{\psi+3\zeta/2}{\sqrt{1+\psi+\chi+\zeta}},\label{xyz}
\ee
so a necessary condition is $\psi+3\zeta/2>0$, which in terms of the original quantities reads 
\be
\frac{4M}{\pi}>\frac{Q^2}{a^2}.
\ee
From (\ref{xyz}) we obtain
\be
\sqrt{1+\psi+\chi+\zeta}=\frac{\zeta}{2}-(1+\chi).\label{tt00}
\ee
Because  $\zeta\leq 0$ and we have the restriction (\ref{cond}), this implies 
\be
\chi<\frac{\zeta}{2}-3<0
\ee
which translates to 
\be
\frac{4\Lambda \alpha}{3}<-\lp\frac{4Q^2\alpha}{a^6}+3\rp.\label{chsim}
\ee
We therefore find that there is no solution to the problem posed if the cosmological constant is positive or null, so the solution to all our demands cannot be within the parameter range of the non exotic configuration found in \cite{yo11}, which has $\Lambda=0$ from the start. Even worse, the latter inequality implies that the metric is ill defined as from a large enough radius it includes an imaginary part. Thus the only possible solution to satisfy exactly all the desired conditions is to examine if they can be fulfilled by a radius $a$ smaller than that of the singular surface $r=r_S$ where the expression inside the square root in Eq. (\ref{metric}) is equal to zero, and to match the region including the throat to two suitable exterior  submanifolds at a radius $b$ such that $a<b<r_S$. As noted before, the charge worsens the situation with the energy condition $\sigma\geq 0$, and as we found just above in (\ref{chsim}) it does not help  in what regards the tide condition ${\cal T}_r=0$. Thus here we assume $Q=0$ so (\ref{chsim}) reduces to $4\Lambda\alpha/3< -3$, and the singular surface radius is much easier to be calculated; the result is  
\be
r_S^4=-\frac{48M\alpha}{\pi\lp 3+4\Lambda\alpha \rp}\label{rsing}
\ee
(which is positive because $4\Lambda\alpha< -9$).

 The throat radius such that ${\cal T}_r=0$ solves the equation (\ref{tt00}) with $\zeta=0$, and is given by
 \be
 a^4  = \frac{36M}{\pi\Lambda\lp3+ 4\Lambda\alpha \rp}\label{athroat}
\ee
(which, again, is positive because both $\Lambda$ and $4\Lambda\alpha +3$ are negative). Then we have 
\be
\frac{r_S^4}{a^4}=-\frac{4\Lambda\alpha}{3}>3,
\ee 
and the throat radius is effectively in the range where the metric is well defined, that is below the singular surface radius. Following Eq. (24), the inequality $\sigma \geq 0$ finally reads
\be
a^4\left(\frac{4\Lambda\alpha}{3}-3\right)-32\alpha a^2  \geq  16\alpha\left(4\alpha-\frac{M}{\pi}\right),\label{a42}
\ee
so, because $\alpha>0$ and $-\Lambda\alpha>0$, it imposes the condition  $M>4\pi\alpha$. Introducing the zero radial crossing tide solution, given by Eq. (\ref{athroat}), this inequality turns into
\be
\frac{3M}{\pi\Lambda}\left(\frac{9-4\Lambda\alpha}{3+4\Lambda\alpha}\right)+48\alpha\sqrt{\frac{M}{\pi\Lambda\lp3+4\Lambda\alpha\rp}}\leq 4\alpha\lp\frac{M}{\pi}-4\alpha\rp.\label{mu1}
\ee
 Note that the requirement of  a negative cosmological constant comes from demanding  a null troublesome crossing contribution to the radial tide. The energy conditions could be fulfilled with just a suitable combination of the mass and the Gauss--Bonnet constant \cite{yo11}. If the GR branch of the geometry had been selected, instead, the possibility of normal matter could only be accomplished in the $\alpha<0$ case \cite{risi08}.
 
We can rearrange (\ref{mu1})  and introduce the function ${\cal F}(M)$ as
\be
{\cal F}(M)\equiv\frac{3M}{\pi\Lambda}\left(\frac{9-4\Lambda\alpha}{3+4\Lambda\alpha}-\frac{4\Lambda\alpha}{3}\right)+48\alpha\sqrt{\frac{M}{\pi\Lambda\lp3+4\Lambda\alpha\rp}}+16\alpha^2, \label{mu11}
\ee
which must be negative or zero.  The restriction $-4\alpha\Lambda>9$ already found renders positive the quantity within the parenthesis of the first term above, and --because $\Lambda<0$-- this allows for the possibility of a solution. A little algebra then serves to find a positive root of the function ${\cal F}(M)$ beyond which all values of the mass satisfy the inequality ${\cal F}(M)\leq 0$. 
For given values of the Gauss--Bonnet parameter $\alpha>0$ and of the cosmological constant $\Lambda<0$, we obtain that the mass $M$ must fulfil the inequality
\be
\sqrt{M}\geq-\frac{4\alpha\sqrt{\pi\Lambda(3+4\alpha\Lambda)}}{9+4\alpha\Lambda}\label{all}
\ee
in order to achieve a thin-shell wormhole supported by ordinary matter (that is, satisfying the energy conditions) and with no troublesome crossing radial tides coming from the singular nature of the matter distribution. Because $-4\alpha\Lambda>9$, the condition (\ref{all}) is always more constraining than the restriction $M>4\pi\alpha$ above. 

\subsection{Avoiding the ill-behaved metric}

The mathematical construction of the preceding section provides a solution to all the demands on the wormhole matter and tides at the crossing surface; however, as it stands, it behaves badly   at both sides outside the throat, because the condition $4\alpha\Lambda/3<-3$ implies the existence of a singular surface beyond which the metric is not real anymore. As we mentioned, then, we should cure this ill behaviour by removing the outer part of each side of the geometry, and pasting them to two submanifolds with well behaved metrics in the range $r\geq b$ with $a<b<r_S$. This can be done in several ways; here we will only propose two simple solutions, each one with different desirable features regarding the matter content or the far behaviour of the geometry. 

In the mathematical construction of the complete space-time, the inner ($a<r<b<r_S$) geometry is given by equations (\ref{metric0}) and (\ref{metric}), while the outer ($b<r$) geometry is defined by
\begin{equation}
{ds}^{2}=-h(r)\,d{\tilde t}^{2}+h^{-1}(r)\,dr^{2}+r^{2}d\Omega^{2}_{3}\label{metric00}
\end{equation}
where
\begin{equation}
h(r)=1+\frac{r^{2}}{4\alpha}\left[1\mp\sqrt{1+\frac{16m\alpha}{\pi r^{4}}-\frac{8q^{2}\alpha}{3 r^{6}}+\frac{4 \lambda \alpha}{3}}\right]\label{metric11},
\end{equation}
with $m$, $q$ and $\lambda$ the exterior mass, charge and cosmological constant respectively. The continuity of the line element imposes  that the time coordinates at each side are related by $f(b)dt^2=h(b)d{\tilde t}^2$. The jump of the extrinsic curvature at each of the outer shells implies a surface energy density given by
\begin{equation}
16\pi b\,\sigma(b)=-\sqrt{h(b)}\left[ 6-\frac{4\alpha}{b^2}\left(2h(b)-6\right)\right]+\sqrt{f(b)}\left[ 6-\frac{4\alpha}{b^2}\left(2f(b)-6\right)\right].\label{sigmaR}
\end{equation}
The pressure, on the other hand, is given by 
 \begin{eqnarray}
16\pi b\,p(b) & = & \sqrt{h(b)}\left[ 4+\frac{bh^{'}(b)}{h(b)}+4\alpha  \frac{ h^{'}(b)}{bh(b)}\left(1-h(b)\right)\right]\nonumber\\
& & -\sqrt{f(b)}\left[ 4+\frac{bf^{'}(b)}{f(b)}+4\alpha  \frac{ f^{'}(b)}{bf(b)}\left(1-f(b)\right)\right].\label{pe2}
\end{eqnarray}
Starting from these general expressions we will exhibit two possible solutions to the problem of the imaginary square root in the metric. 

\begin{enumerate}[(i)]

\item 
The easiest possibility to match the inner geometry to an outer well behaved one is to accept $\sigma(b)=0$ and examine if this can be fulfilled for $f(b)=h(b)$ with $b$ a radius larger than $a$ and smaller than $r_S$. This immediately rules out outer metrics of the GR branch form, and we look for the simplest case in which  $q=\lambda=0$.  Thus we need $f(b;M,Q=0,\Lambda<0)=h(b;m,q=0,\lambda=0)$, so we have
\be
\sqrt{1+\frac{16M\alpha}{\pi b^4}+\frac{4\Lambda\alpha}{3}}=\sqrt{1+\frac{16m\alpha}{\pi b^4}}
\ee
which can only be satisfied if $m<M$. This condition is solved by
\be
b^4=\frac{12M}{\pi\Lambda}\lp \frac{m}{M}-1\rp\label{Rshell}
\ee
and we must check if such radius lies within $a$
and $r_S$. To ensure $b>a$, according to (\ref{Rshell}) and (\ref{athroat}) this implies the condition
\be
\frac{m}{M}<\frac{6+4\Lambda\alpha}{3+4\Lambda\alpha}<1.
\ee
The second inequality above is ensured by the restriction $4\Lambda\alpha<-9$. On the other hand, $b<r_S$ would imply, according to (\ref{rsing}), that $m/M>3/(3+4\Lambda\alpha)$, which under the conditions demanded is always satisfied for positive masses. 
 
Of course, the nature of matter on the additional outer shells is to be addressed. Working under the condition $f(b)=h(b)$ this simplifies considerably; after some algebra we obtain 
\be
p(b)=\frac{m-M}{2\pi^2 b^3\sqrt{f(b)}}
\ee
which is negative because $m<M$. Thus the matter in the outer shells  would not fulfil the weak energy condition. However, in the context of wormholes and other peculiar configurations it was proposed to characterize the exotic matter by means of a volume quantifier $\Omega$ based on an average in the direction in which an observer moves to go across the distribution \cite{viskardad,bavis,nzhk04}. Thus the most usual choice has become the integral including the energy density $\rho$ and the radial pressure (see \cite{nzhk04} and also \cite{grg06}):
\be
\Omega =\int (\rho+p_{r}) \sqrt{-g}\, d^4x
\ee
where $g$ is the determinant of the metric tensor. Because each infinitely thin shell placed at $r=b$ does not exert radial pressure and the energy density is located on a surface, so that at each side $
\rho =\delta({r-b})\sigma(b)
$ (with $\delta$ the Dirac delta distribution), then for each outer shell in our construction we simply have 
\be
\Omega = \int\limits_{0}^{2\pi}\int\limits_{0}^{\pi}\int\limits_{0}^{\pi} \sigma(b) \left.
\sqrt{-g}\right|_{r=b}\,d\theta_1\,d\theta_2\, d\theta_3=2\pi^2 b^3\sigma(b) =0.
\ee
With the definition adopted, the amount of exotic matter associated to the whole configuration would then be zero.  Moreover, if the repulsive character of the force on rest particles associated to the ``wrong'' sign of the mass term in the asymptotic limit of the non-GR branch is to be avoided (see below), this can be done by the simple choice $m=0$; this does not contradict the conditions required for the definition of the shells at $r=b$, but only fixes their radii at $b=\lp-12M/(\pi\Lambda)\rp^{1/4}$.

\item

We can again assume a simple situation in which both the inner and the outer metrics are of the form of the non-GR branch, and $q=\lambda=0$.  Now, however, no forced link between the inner and outer geometries is established from the start. This will allow to prove the possibility of a construction solving the problem posed by the square root in $f_+(r)$, and with matter satisfying the energy conditions everywhere. Analogously as in (\ref{defs}), we define 
 \be\label{defs2}
\frac{16M\alpha}{\pi b^{4}}\equiv \omega,\ \ \  \ \ \frac{4 \Lambda \alpha}{3}\equiv \chi, \ \ \ \  \ \frac{16m\alpha}{\pi b^{4}}\equiv \phi .
\ee
With these definitions and after some algebra we obtain
\begin{eqnarray}
16\pi b\,\sigma(b) & = & 2\sqrt{1+\phi}\sqrt{1+\frac{b^2}{4\alpha}\lp 1+\sqrt{1+\phi}\rp}\nonumber\\
& & -\,2\sqrt{1+\omega+\chi}\sqrt{1+\frac{b^2}{4\alpha}\lp 1+\sqrt{1+\omega+\chi}\rp}\nonumber\\
& & +\, 4\lp 1+\frac{4\alpha}{b^2}\rp\sqrt{1+\frac{b^2}{4\alpha}\lp 1+\sqrt{1+\omega+\chi}\rp}\nonumber\\
& & -\,4\lp 1+\frac{4\alpha}{b^2}\rp\sqrt{1+\frac{b^2}{4\alpha}\lp 1+\sqrt{1+\phi}\rp}.\label{sigmaRR}
 \end{eqnarray}
 The sum of the energy density and the pressure is given by
 \begin{eqnarray}
16\pi b\,\lp\sigma(b)+p(b)\rp & = & 2\sqrt{1+\phi}\sqrt{1+\frac{b^2}{4\alpha}\lp 1+\sqrt{1+\phi}\rp}\nonumber\\
& & -\,2\sqrt{1+\omega+\chi}\sqrt{1+\frac{b^2}{4\alpha}\lp 1+\sqrt{1+\omega+\chi}\rp}\nonumber\\
& &+\,\frac{16\alpha}{b^2}\sqrt{1+\frac{b^2}{4\alpha}\lp 1+\sqrt{1+\omega+\chi}\rp}\nonumber\\
& & -\,\frac{16\alpha}{b^2}\sqrt{1+\frac{b^2}{4\alpha}\lp 1+\sqrt{1+\phi}\rp}\nonumber\\
& &  +\,\frac{b^2\lp 1+\chi+\sqrt{1+\omega+\chi}\rp}{2\alpha\sqrt{1+\frac{b^2}{4\alpha}\lp 1+\sqrt{1+\omega+\chi}\rp}}\nonumber\\ 
& &-\,\frac{b^2\lp 1+\sqrt{1+\phi}\rp}{2\alpha\sqrt{1+\frac{b^2}{4\alpha}\lp 1+\sqrt{1+\phi}\rp}}
.\label{s+p2}
 \end{eqnarray}
 Because $a<b<r_S$ with $a$ and $r_S$ given by (\ref{rsing}) and (\ref{athroat}) we know that $0<\sqrt{1+\omega+\chi}<2$. On the other hand,  in the expressions above the highest power of the root $\sqrt{1+\phi}>1$ is that of the positive contributions  to $\sigma(b)$ and to $\sigma(b) +p(b)$ associated to the outer parameters (first term in both (\ref{sigmaRR}) and (\ref{s+p2})).   We are free to set the quantity $\phi$ (and consequently the root $\sqrt{1+\phi}$) as large as desired by simply choosing a large enough outer mass $m$. Hence we can ensure that for a given set of parameters $M$, $\Lambda$ and $\alpha$, for any $b$ in the range between the throat radius $a$ and the singular radius $r_S$, the mass of the outer geometries can be chosen so that the weak energy condition is fulfilled at each exterior shell.
\end{enumerate}
In what regards the behaviour of the outer regions $r>b$ at each side of the wormhole throat in both solutions adopted,  these would not be asymptotically flat, its asymptotic form given by 
 \be
 h(r) \to  1+\frac{2m}{\pi r^2}+\frac{r^2}{2\alpha}\label{ametrice1} .
\ee
However, because for positive $\alpha$ the effective cosmological constant that we can read from this expression is negative, the positive feature of no cosmological horizons would be achieved in both proposed constructions.

\section{Summary}

We have shown the possibility of spherical thin-shell wormhole geometries supported by ordinary matter and with no troublesome contributions to crossing radial tides associated to the singular character of the Riemann tensor at the throat. This is enabled by Gauss--Bonnet terms, as pure Einstein gravity forces compact wormholes to be threaded by  exotic matter. Given known previous results, we focused our study, mainly, on  positive values of the Gauss--Bonnet parameter and the non-GR branch of Wilstshire solution. We not only have succeeded in our primary aim; besides, singularities and regions with ill-defined metric related with the behaviour of the branch selected have been avoided by a suitable secondary cut and paste procedure in which  the region including the throat is matched to two outer regions with an admissible (at least formally) behaviour.  We have presented two simple ways to achieve this, with the resulting complete geometry corresponding to an inner region associated to a mass $M$, a negative cosmological constant $\Lambda$ and zero charge, and two exterior regions associated to a mass $m$; the latter should satisfy different conditions in each case. In the first proposed solution, the total amount of exotic matter of the whole construction is zero, and the repulsive far behaviour of the metric can be avoided. In the second one we have demonstrated the possibility of well behaved exterior metrics by placing outer shells constituted by normal matter, that is shells which satisfy the weak energy condition. Being our central aims the avoidance of exotic matter and troublesome radial tides at the wormhole throat, we have not explored other possible outer geometries matched to the inner region. For example, the possibility of asymptotically flat metrics working with the GR branch as the exterior metric could be considered, as well as  tides in the vicinity of the outer shells; however, according to the results in \cite{nos23,nos25}, the latter should not constitute a serious issue regarding traversability.

\section*{Acknowledgments}

This research was funded by CONICET and UBA.

\end{document}